\documentstyle[twocolumn,seceq,psfig]{jpsj}

\def\runtitle{NMR and Neutron Scattering Studies of CuV$_2$O$_6$}
\def\runauthor{Jun {\sc Kikuchi} \it et al.}

\title
{
NMR and Neutron Scattering Studies of Quasi One-Dimensional Magnet
CuV$_2$O$_6$ }

\author
{ 
Jun {\sc Kikuchi}\footnote{E-mail: kikuchi@ph.noda.sut.ac.jp},
Kazuhiro {\sc Ishiguchi},
Kiyoichiro {\sc Motoya}, Masayuki {\sc Itoh}$^{1}$,\\
Kazunori {\sc Inari}$^{2}$, Naotoshi {\sc Eguchi}$^{2}$ and Jun {\sc
Akimitsu}$^{2}$ }

\inst
{
Department of Physics, Faculty of Science and Technology, Science
University of Tokyo,\\
2641 Yamazaki, Noda, Chiba 278-8510\\
$^1$Department of Physics, Faculty of Science, Chiba University,\\
1-33 Yayoi-cho, Inage-ku, Chiba 263-8522\\
$^2$Department of Physics, College of Science and Engineering, Aoyama-
Gakuin University,\\
6-16-1 Chitosedai, Setagaya-ku, Tokyo 157-8572
}

\recdate
{
April 17, 2000
}

\abst
{
Magnetic properties of a quasi one-dimensional magnet CuV$_2$O$_6$ have
been studied by means of susceptibility measurement, NMR and powder neutron
diffraction.  Cu$^{2+}$ spins ($S=1/2$) order antiferromagnetically below
22.6 K, and the saturation moment is about 0.70 $\mu_{\rm B}$ at 0 K.
The magnetic structure determined in the present study indicates that in
CuV$_2$O$_6$ one-dimensional antiferromagnetic spin chains are along
[110], suggesting a dominant antiferromagnetic exchange interaction
between the next-nearest-neighbor spins rather than the
nearest-neighbor ones.  }

\kword
{CuV$_2$O$_6$, NMR, neutron diffraction, low-dimensional magnet}

\begin{document}
\sloppy
\maketitle

\section{Introduction}
A Heisenberg antiferromagnetic linear chain (HAFC) with $S=1/2$ has attracted
much attention because of its critical nature of the ground state.  Quantum
fluctuations in the $S=1/2$ HAFC are so strong that the ground state is
disordered and is gapless, and the spin-spin correlation function decays 
as a power law in an infinite correlation length.  Because of its
criticality, a small perturbation causes a drastic change of the
ground state to the one with spontaneous broken symmetry.  A coupling
with phonons, for example, results in a new disordered ground state
with an energy gap for the excitation channels, accompanying
spontaneous dimerization of the spins.  This is known as the
spin-Peierls transition.\cite{jacobs76} An interchain coupling is also
a trigger for transition to the three-dimensionally-ordered ground
state with spontaneous sublattice magnetization, which is usually
encountered in a lot of ``quasi'' one-dimensional materials.

\begin{figure}
\centerline{\psfig{figure=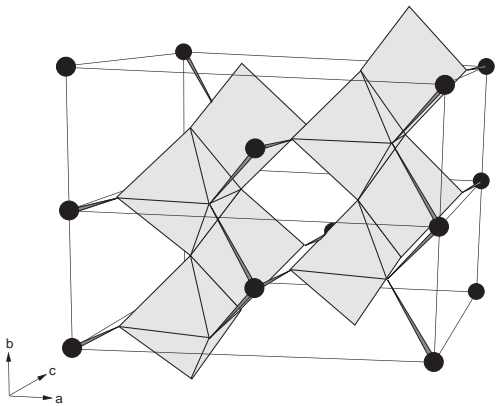,width=6.5cm,angle=0}}
\medskip 
\caption{Crystal structure of CuV$_2$O$_6$.  The solid circles are Cu atoms
and the gray polyhedra represent VO$_5$ pyramids.  Cu-O bonds are
shown for oxygens shared by CuO$_6$ and VO$_5$ polyhedra.  }
\label{structure}
\end{figure}

An insulating copper vanadate CuV$_2$O$_6$ is one of the possible candidates 
of an $S=1/2$ HAFC. CuV$_2$O$_6$ crystallizes in the triclinic
structure of space group $P\overline{1}$ with lattice parameters $a =
9.168$ {\AA}, $b = 3.543$ {\AA}, $c = 6.478$ {\AA}, $\alpha =
92.25^{\circ}$, $\beta = 110.34^{\circ}$ and $\gamma =
91.88^{\circ}$.\cite{calvo73,fnSG} The structure consists of linear
chains of edge-sharing CuO$_6$ octahedra along [010] or the $b$ axis,
bridged by VO$_5$ pyramids forming a zig-zag chain (or also viewed as
double chains) of V atoms running along [010] as well (see Fig.\
\ref{structure}).  Valences of the ingredient atoms are Cu$^{2+}$,
V$^{5+}$ and O$^{2-}$, and the Cu$^{2+}$ ion has a spin one half as only
the magnetic constituent.  The Cu-Cu distances are 3.543 {\AA} along
[010], 4.860 {\AA} along [110], 4.968 {\AA} along [1$\overline{1}$0]
and 6.478 {\AA} along [001].  Because the distance between Cu atoms is
the shortest along [010] and there seems to be a relevant exchange
path via the shared CuO$_6$ octahedral edge in that direction, one
naively expects that CuV$_2$O$_6$ has one-dimensional (1D) chains of
Cu$^{2+}$ spins along [010] and that the magnetic properties can well
be described in terms of an $S=1/2$ HAFC. Indeed, a broad maximum of
the magnetic susceptibility has been observed around 44
K,\cite{vasilev99} which indicates strong short-range
antiferromagnetic correlation between Cu$^{2+}$ spins.  On the other
hand, it has been reported from the AFMR measurement that CuV$_2$O$_6$
has a long-range-ordered ground state below 23 K,\cite{vasilev99}
implying a relatively large interchain coupling.

Although the low-dimensional character of the spin system in CuV$_2$O$_6$ is
evident from the preceding work,\cite{vasilev99} it seems that the
magnetic properties of CuV$_2$O$_6$ have not yet been well
characterized because of the lack of microscopic experiments.  It is
therefore worthwhile to conduct an experiment on CuV$_2$O$_6$ such as
the nuclear magnetic resonance (NMR) and the neutron diffraction.  In
this paper, we present results of magnetic susceptibility, NMR and
powder neutron diffraction measurements on CuV$_2$O$_6$.  We confirmed that
CuV$_2$O$_6$ exhibits commensurate antiferromagnetic ordering below
the N\'{e}el temperature of 22.6 K. We unexpectedly found that the
spin chains are ordered antiferromagnetically not along [010] but [110],
corresponding to the direction of next-nearest neighbors for Cu, the
origin of which will be argued based on the local electronic state of
the Cu$^{2+}$ ion.

\section{Experiments}
\label{sec:experiments}
Polycrystalline samples of CuV$_2$O$_6$ were prepared by the
solid-state reaction of CuO and V$_2$O$_5$.  They were mixed in an
alumina crucible and fired in air at 500 $^\circ$C. The firing and
intermediate grinding were repeated until the impurity phases such as
CuV$_2$O$_7$ were not detected by the powder X-ray diffraction
measurement.  The time in obtaining a single phase was totally of
about 200 hours.  Magnetic susceptibility was measured using a
SQUID magnetometer (Quantum Design MPMS-5s) at 1.5 T. $^{51}$V NMR spectra
were taken by recording the spin-echo signals with a Box-car
integrator while sweeping the external magnetic field at a fixed
frequency.  The Cu NMR spectrum was taken by recording the spin-echo
signals under zero external field point by point of frequencies. 
Powder neutron-diffraction measurements were performed on the
triple-axis spectrometer (T1-1) in a two-axis mode installed at JRR-3M
in JAERI at Tokai.  Pyrolytic graphite monochromator and filter were
used to select the incident neutrons.  The wave number of incident
neutrons was 2.575 {\AA$^{-1}$} and the horizontal beam collimation,
7$^{\prime}$-40$^{\prime}$-40$^{\prime}$-40$^{\prime}$, was used in
the experiments.

\section{Results and Analysis}
\label{sec:results}
\subsection{Magnetic Susceptibility}
\begin{figure}[t]
\centerline{\psfig{figure=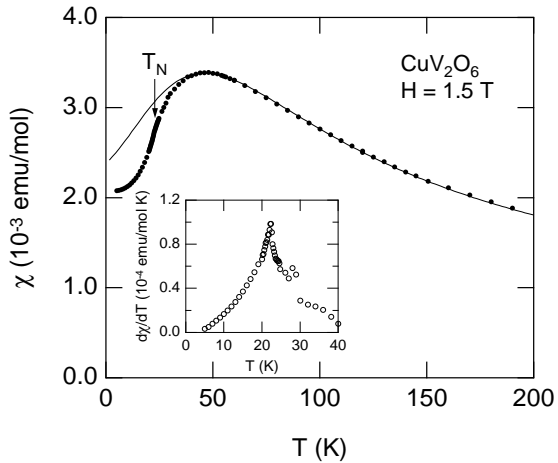,width=8.0cm,angle=0}}
\medskip 
\caption{Temperature dependence of the magnetic susceptibility of
CuV$_2$O$_6$.  N\'{e}el temperature is indicated by an arrow.  The solid line
is a fit of the data above 40 K to the susceptibility of an $S=1/2$
Heisenberg antiferromagnetic linear chain.  The inset shows $d\chi/dT$
which exhibits a sharp peak at $T_{\rm N}$.}
\label{susceptibility}
\end{figure}

The temperature dependence of the magnetic susceptibility of
CuV$_2$O$_6$ is shown in Fig.\ \ref{susceptibility}.  The
susceptibility exhibits a broad maximum around $T_{\rm max} \approx
48$ K which is characteristic of low-dimensional system with
antiferromagnetic interactions.  Above $T _{\rm max}$, the
susceptibility can well be fitted to the Bonner-Fisher curve for the $S=1/2$
HAFC.\cite{bonner64} In order to estimate the intrachain exchange
coupling $J$ between Cu$^{2+}$ spins, we fitted the temperature ($T$)
dependence of the susceptibility $\chi$ to the following
formula;\cite{hatfield81}
\begin{eqnarray}
	\chi = &&\chi_{0}
	+ {{N_{\rm A}\,{g^2}\,{\mu^2_{\rm B}}}\over {k_{\rm B}\,T}} \nonumber \\
	&&\times 
	{{0.25 + 0.14995\,x +0.30094\,{x^2}}\over {1 + 1.9862\,x +
	0.68854\,{x^2} + 6.0626\,{x^3}}}.
\end{eqnarray}

\noindent
Here $\chi_{0}$ is a $T$ independent part of $\chi$, $N_{\rm A}$ is
Avogadro's number, $\mu_{\rm B}$ is the Bohr magneton and $x=J/k_{\rm
B}T$.  From the least-squares fit of the data above 40 K, we obtained
$\chi_{0}= 3.5 \times 10^{-4}$ emu/mol, $g$ = 2.0 and $J/k_{\rm B}$ =
36 K. A relatively large value of $\chi_{0}$ may be attributed to
Van-Vleck orbital paramagnetism of Cu$^{2+}$ ions.  As the temperature
is decreased below $T _{\rm max}$, the observed $\chi$ begins to
deviate from the Bonner-Fisher curve.  This is considered to be a
crossover to higher dimensions resulting from the interchain coupling. 
Indeed, long-range antiferromagnetic ordering takes place below 22.6 K
as will be shown below.

Although we cannot see a clear anomaly in the $\chi -T$ curve
indicative of long-range magnetic ordering, we observed a
sharp peak in the temperature derivative of the susceptibility
$d\chi/dT$ at 22.4 K as shown in the inset of Fig.\
\ref{susceptibility}.  It is reasonable to consider that the anomaly is
due to the long-range magnetic ordering of Cu$^{2+}$ spins.

\subsection{$^{51}$V NMR}
\begin{figure}
\centerline{\psfig{figure=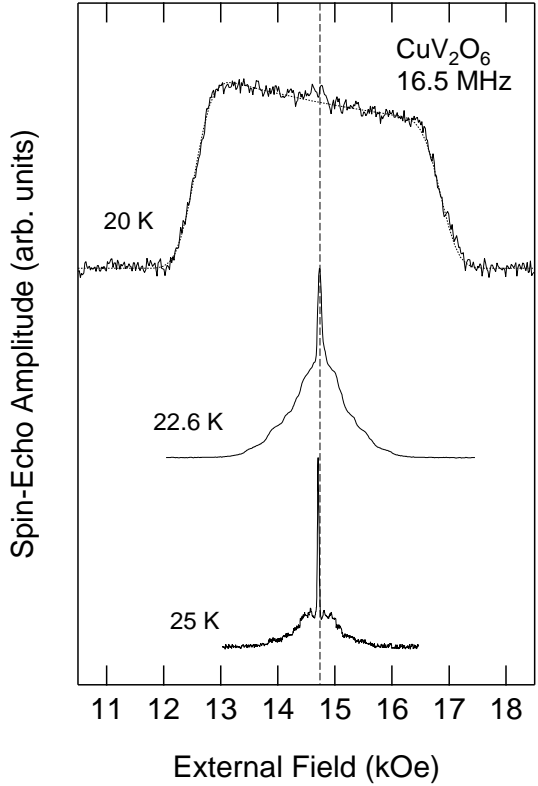,width=9.0cm,angle=0}}
\medskip 
\caption{Temperature variation of the field-swept NMR spectrum of
$^{51}$V in CuV$_2$O$_6$ taken at 16.5 MHz.  The dashed line indicates
a position of zero shift for $^{51}$V. The dotted line is a fit of the
experimental data at 20 K to eq.\ (\ref{AFspectrum}) broadened by an
inhomogeneous distribution of the internal field.}
\label{51Vspectra}
\end{figure}

Figure \ref{51Vspectra} shows the field-swept NMR spectrum at various
temperatures taken at an NMR frequency of 16.5 MHz.  The spectra
exhibit powder patterns which distribute around the zero-shift
position for $^{51}$V nuclei.  The fine structure in the spectrum
observed above 22.6 K can be assigned as singularities resulting from
the electric quadrupole interaction for nuclei with a nuclear spin $I =
7/2$.  We therefore conclude that the observed signal comes from
$^{51}$V nuclei ($I = 7/2$) on the nonmagnetic V site.

The spectrum starts to be broadened
suddenly below 22.6 K to have a rectangular shape with the center of
gravity near the unshifted resonance position for $^{51}$V. The
broadening evidences long-range magnetic ordering of Cu$^{2+}$ spins
giving rise to a finite internal magnetic field at the V site.  The
rectangular shape of the spectrum is typical for nonmagnetic nuclear
sites in the antiferromagnet with the internal field much smaller than
the applied one.  The line shape also indicates a commensurate
antiferromagnetic structure in which all the V sites feel a unique
internal-field strength from Cu$^{2+}$ spins.

In order to determine the magnitude of the internal field $H_{\rm n}$ at the V
site which is proportional to the sublattice magnetization,\cite{jaccarino65} we
calculated the line shape of the NMR spectrum in the antiferromagnetic
state and fitted the observed spectrum to the calculated one. 
Details of the derivation and the fitting procedure should be referred to the
appendix.  We describe here an essence of the calculation.  For a
powder sample in which a direction of the internal field is randomly
distributed with respect to that of the applied field $H$, the NMR spectrum
$f(H)$ as a function of $H$ has the following form;
\begin{equation}
\label{AFspectrum}
	f(H) \propto {{H^2-H_{\rm n}^2+\omega^2 /\gamma^2}\over {H_{\rm n} H^2}}.
\end{equation}

\noindent 
Here $\omega$ is an NMR frequency which is assumed to be larger than
$\gamma H_{\rm n}$ ($\gamma$ being the nuclear gyromagnetic ratio). 
The spectrum has two cutoff fields, $\omega /\gamma + H_{\rm n}$ and
$\omega /\gamma - H_{\rm n}$, resulting in the sharp edges in the
spectrum, which are usually smeared by an inhomogeneous distribution
of the internal field present in a real crystal.  We take account of
this effect by assuming a gaussian distribution function for $H_{\rm
n}$.  The result of the fit is shown in Fig.\ \ref{51Vspectra} by the
dotted line.  The quality of the fit is quite good and the magnitude
of the internal field at 20 K was determined as $H_{\rm n} = 2.15 \pm
0.06$ kOe.

\begin{figure}[t]
\centerline{\psfig{figure=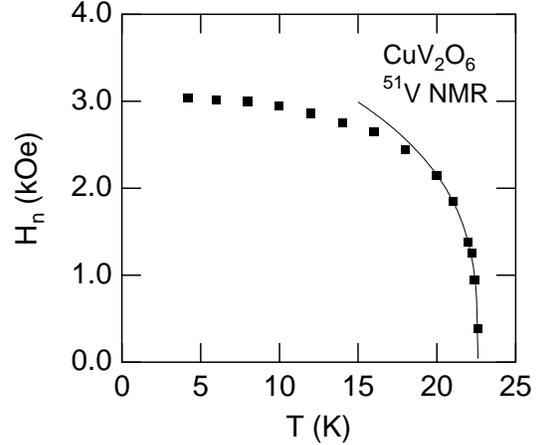,width=8.0cm,angle=0}}
\medskip 
\caption{Temperature dependence of the internal magnetic field at the 
V site in CuV$_2$O$_6$.  The solid line is a fit of the data above 20 K
to the form $H_{\rm n} \propto (1-T/T_{\rm N})^{\beta}$ with $T_{\rm
N} = 22.60$ K and $\beta = 0.31$.}
\label{MsNMR}
\end{figure}

The $T$ dependence of the internal field at the V site is shown in Fig.\
\ref{MsNMR}.  The internal field becomes nonzero below about 23 K and
saturates to a value of about 3.1 kOe at $T$ = 0 K. The N\'{e}el
temperature $T_{\rm N}$ was determined by fitting $H_{\rm n}$ to be
proportional to $(1-T/T_{\rm N})^{\beta}$ near $T_{\rm N}$, where
$\beta$ is the critical exponent for sublattice magnetization.  From the
least-squares fit of the data above 20 K, we obtained $T_{\rm N} =
22.60 \pm 0.05$ K and $\beta = 0.31 \pm 0.01$.  The exponent is close
to those for three-dimensional (3D) magnetic systems,\cite{jongh74}
suggesting the 3D nature of magnetic ordering.

\begin{figure}[t]
\centerline{\psfig{figure=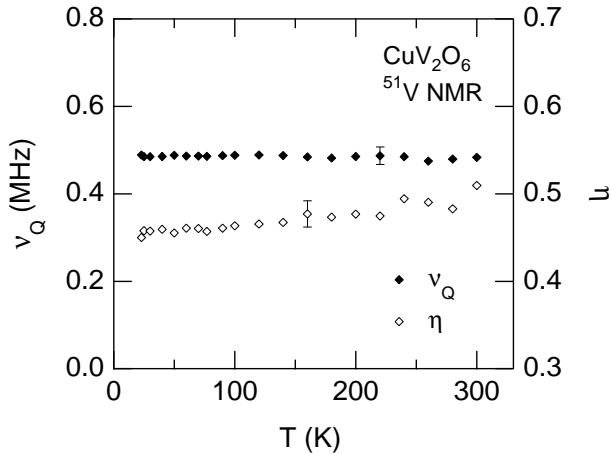,width=8.5cm,angle=0}}
\medskip 
\caption{Temperature dependence of the nuclear quadrupolar frequency
$\nu_{\rm Q}$ and the asymmetric parameter $\eta$ at the V site in
CuV$_2$O$_6$.  Error bars are shown for some data points.}
\label{51EFG}
\end{figure}

\begin{figure}
\centerline{\psfig{figure=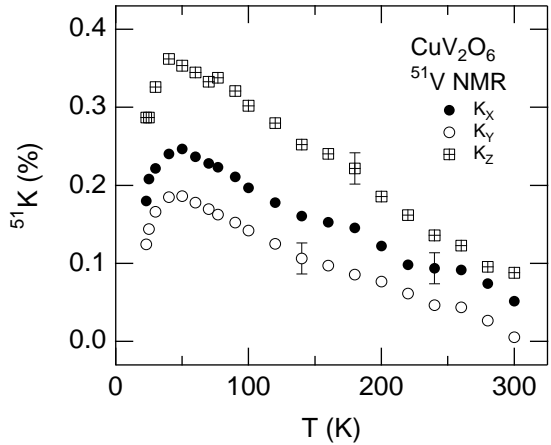,width=8.5cm,angle=0}}
\medskip 
\caption{Temperature dependence of the NMR shift at the V site in
CuV$_2$O$_6$.  Error bars are shown for some data points.}
\label{51KvsT}
\end{figure}

In the paramagnetic state above $T_{\rm N}$, the $^{51}$V NMR
spectrum exhibits a powder pattern resulting from the anisotropy of NMR shift
and electric-field gradient (EFG) at the V sites.  All the peaks and
shoulders observed in the satellites can be assigned as singularities
which appear when the external magnetic field is parallel to one of the
principal axes of the EFG tensor.  By analyzing the positions of peaks and
shoulders in the satellites, one can determine the tensor components
of the NMR shift, and the EFG parameters such as the nuclear
quadrupolar frequency $\nu_{\rm Q}$ and the asymmetric parameter
$\eta$.\cite{abragam61} Taking account of the quadrupolar effect on
the Zeeman interaction as a perturbation up to second order, we
determined the NMR shifts and the EFG parameters at each temperature. 
The results are shown in Figs.\ \ref{51EFG} and \ref{51KvsT}.\\
\indent
Due to low point symmetry, the EFG at the V sites is rather
asymmetric with an $\eta$ value of $0.45-0.51$ which depends weakly on
temperature.  As a V$^{5+}$ ion has no valence electron, the EFG at
the V site is mainly determined by the contribution from surrounding
lattice ions which may be regarded as point charges.  The lattice
contribution to the EFG evaluated in terms of the point-charge model
yields $\eta$ of 0.51 at room temperature, which agrees well with the
experiment.  This suggests that the large asymmetry of the EFG at the
V site results from the asymmetric configuration of surrounding
lattice ions.\linebreak
\begin{figure}
\centerline{\psfig{figure=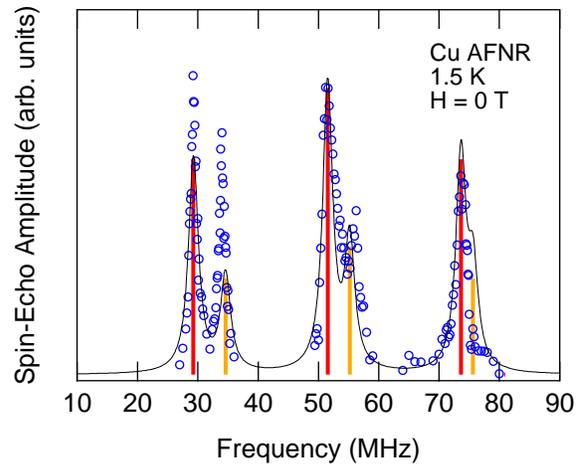,width=9.0cm,angle=0}}
\medskip 
\caption{The zero-field NMR spectrum of Cu in CuV$_2$O$_6$ taken at
1.5 K. The data points are shown by open circles.  The vertical lines
indicate resonance frequencies and intensities for $^{63}$Cu (black)
and $^{65}$Cu (gray) nuclei calculated using parameters; $H_{\rm n}$ =
4.56 T, $^{63}\nu_{\rm Q}$ = 22.2 MHz and $\eta$ = 0.32.  The solid
line is a simulation of the NMR spectrum based on the calculated
resonance frequencies and intensities broadened by a lorentzian width
of 1.0 MHz.}
\label{CuAFNR}
\end{figure}
\indent
The NMR shifts $K_{\mu}$ ($\mu$ = $X$, $Y$ and $Z$) exhibit a
broad maximum around 50 K, corresponding to the bulk $\chi$ as shown
in Fig.\ \ref{51KvsT}.  A standard $K-\chi$ analysis indicated that
$K_{\mu}$'s are proportional to $\chi$.  From the proportionality
constants between $K_{\mu}$'s and $\chi$, we determined the tensor
components of the hyperfine interaction at the V site along the $\mu$
principal axes of the EFG as $A_{X} = 4.7 \pm 0.2~{\rm kOe}/\mu_{\rm
B}$, $A_{Y} = 4.3 \pm 0.2~{\rm kOe}/\mu_{\rm B}$ and $A_{Z} = 7.0 \pm
0.3~ {\rm kOe}/\mu_{\rm B}$.  $A_{\mu}$'s are relatively large for
nuclei on the nonmagnetic site which indicates finite spin transfer
from neighboring Cu$^{2+}$ spins.  It should also be noted that the
hyperfine tensor is anisotropic.  One of the origins of an anisotropy
in the hyperfine tensor is the classical dipolar field from
surrounding electronic spins, which can be evaluated numerically if we
know the atomic positions of both V and Cu.  Based on the given
lattice parameters,\cite{calvo73} we calculated the classical dipolar
field at the V site along the lattice EFG principal axes to be
$A_{X}^{\rm dip} = 0.26$ kOe/$\mu _{\rm B}$, $A_{Y}^{\rm dip} = -0.67$
kOe/$\mu _{\rm B}$ and $A_{Z}^{\rm dip} = 0.45$ kOe/$\mu _{\rm B}$. 
Since the observed hyperfine coupling constants can be expressed as
$A_{\mu} = A_{\mu}^{\rm tr} + A_{\mu}^{\rm dip}$, the anisotropy of
the transferred hyperfine coupling ($A_{\mu}^{\rm tr}$) can be
obtained by subtracting the dipolar coupling from the observed one. 
We then obtained the following values of the transferred hyperfine
coupling; $A_{X}^{\rm tr} = 4.4 \pm 0.2~{\rm kOe}/\mu_{\rm B}$,
$A_{Y}^{\rm tr} = 5.0 \pm 0.2~{\rm kOe}/\mu_{\rm B}$ and $A_{Z}^{\rm
tr} = 6.5 \pm 0.3~{\rm kOe}/\mu_{\rm B}$.  Although the transferred
hyperfine coupling is dominated by the isotropic part $A_{\rm
iso}^{\rm tr} = {1\over 3}\Sigma_{\mu} A_{\mu}^{\rm tr} = 5.3 \pm
0.7~{\rm kOe}/\mu _{\rm B}$, it has a finite anisotropy not due to the
classical dipolar field.  This suggests that the anisotropic orbital
such as $2p$, $3p$ and $3d$ orbitals on V atoms are polarized by spin
transfer from neighboring Cu atoms.

\subsection{Cu NMR}
We observed an NMR of Cu nuclei under zero external magnetic field
well below $T_{\rm N}$, which again confirms long-range magnetic
ordering.  Figure \ref{CuAFNR} shows the zero-field NMR spectrum of Cu
taken at 1.5 K. No signal is observed except the observed one in a
frequency range from 20 to 100 MHz.  The observed five resonance lines are
signals from the two isotopes, $^{63}$Cu and $^{65}$Cu, which both
have nuclear spin $I$ of 3/2, split by the electric quadrupolar
interaction.  The absence of one of the six lines expected in the
presence of the quadrupolar interaction is due to an accidental
overlap of satellite lines of the two isotopes.
\begin{fullfigure}[t]
\centerline{\psfig{figure=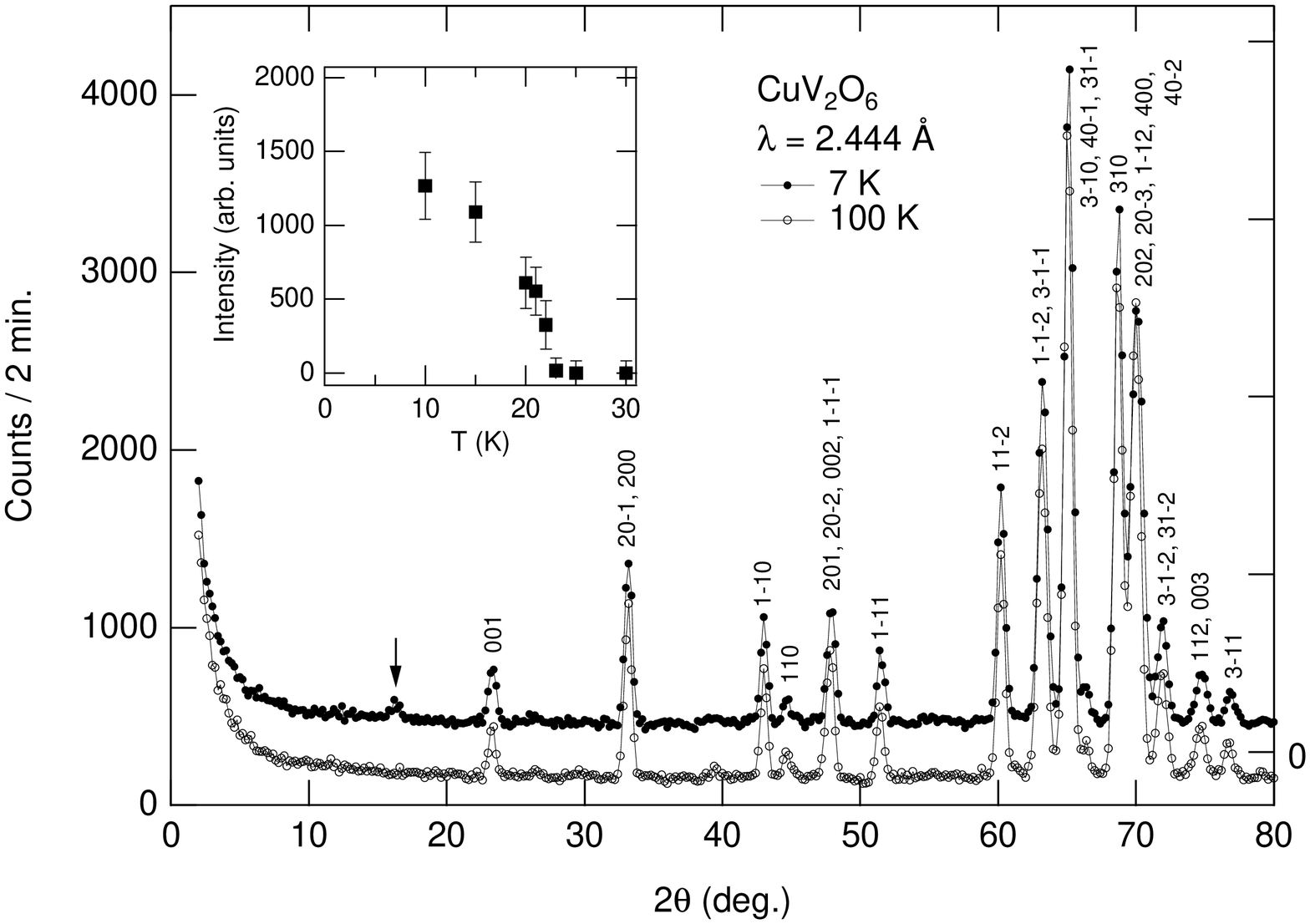,width=16.0cm,angle=0}}
\medskip 
\caption{Powder neutron diffraction patterns from CuV$_2$O$_6$ at 7
(offset by 300 counts) and 100 K. The inset shows temperature dependence of
the integrated intensity of the Bragg peak with the scattering angle
$2\theta$ of 16.3$^\circ$.  An arrow indicates the magnetic Bragg
peak.}
\label{fullscan}
\end{fullfigure}

In order to determine the internal magnetic field $H_{\rm n}$, the
quadrupolar frequency $\nu_{\rm Q}$ and the asymmetric parameter $\eta$
at the Cu site, we calculated energy levels of a nuclear spin with $I
= 3/2$ by solving a secular equation of the nuclear Hamiltonian 
of nuclear Zeeman and electric quadrupole interactions.  This is
because a perturbation theory cannot be applied to calculate energy
levels and resonance frequencies in the present case where the
relative strength of Zeeman and quadrupole interactions is unknown and
may be comparable in magnitude.  The Hamiltonian for Cu nuclei is
generally written as
\begin{eqnarray}
\label{ZeemQuad}
	{\cal H} &=& -\gamma \hbar {\mib I} \cdot {\mib H}_{\rm n} \nonumber \\
	&+&
	{h \nu_{\rm Q}\over 6}[3{I_Z}^2-I(I+1)+{\eta \over 2}({I_+}^2+{I_-}^2)].
\end{eqnarray}

\noindent 
The first and second terms are the nuclear Zeeman and electric quadrupolar
interactions, respectively, and the coordinate refers to the principal axes
of the EFG tensor.  $\gamma$ is the nuclear gyromagnetic ratio
($^{63}\gamma$ = 11.285 MHz/T and $^{65}\gamma$ = 12.089 MHz/T), $\hbar$
is Planck's constant, ${\mib H}_{\rm n}$ is the internal magnetic field
and ${\mib I}$ is the nuclear spin operator.  When a direction of the
internal magnetic field is parallel to one of the principal axes of
the EFG tensor, the Hamiltonian\ (\ref{ZeemQuad}) can be diagonalized
analytically, so that the energy levels and the transition
probabilities between them can be obtained in an analytical form.  On the
assumption that the internal magnetic field is along either $X$, $Y$
or $Z$ axis of the EFG tensor, we tried to reproduce the observed
resonance frequencies by tuning parameters $H_{\rm n} = |{\mib H}_{\rm
n}|$, $^{63}\nu_{\rm Q}$ and $\eta$.\cite{nQ65} A good fit was
obtained when the direction of the internal field is along the $Z$ axis
of the EFG tensor.  The obtained parameters are $H_{\rm n}$ = 4.56 T,
$^{63}\nu_{\rm Q}$ = 22.2 MHz and $\eta$ = 0.32, which were used to
simulate the NMR spectrum shown by the solid line in Fig.\
\ref{CuAFNR}.

\subsection{Powder Neutron Diffraction}
Our NMR experiments presented above confirmed long-range
magnetic ordering of Cu$^{2+}$ spins in CuV$_2$O$_6$.  However, the
magnetic structure can hardly be determined from NMR on powder specimens.  We
therefore performed a powder neutron diffraction measurement on
CuV$_2$O$_6$.

Figure \ref{fullscan} shows the powder diffraction patterns from
CuV$_2$O$_6$ at 7 and 100 K. At 7 K ($< T_{\rm N}$), a new peak
appears at the scattering angle of $2\theta = 16.3^\circ$.  As shown
in the inset of Fig.\ \ref{fullscan}, the peak disappears above
$T_{\rm N}$ and is confirmed to be magnetic in origin.  In table
\ref{NDintensity}, we summarized the scattering angle and intensity of
the observed peaks at 10 K and 30 K along with the Miller indices
based on the chemical unit cell.  Within experimental accuracies,
there is no magnetic reflection except the observed one.
\begin{fulltable}
\caption{The observed Bragg angles and intensities of scattered neutrons
at 10 and 30 K in CuV$_2$O$_6$.  The intensities are normalized to 1000
at the Bragg angle of 65.13$^\circ$.  Errors in the Bragg angles are
within 0.05$^\circ$.  Large errors in $I_{\rm obs}$ are due to
incoherent scattering of neutrons by vanadium.}
\label{NDintensity}
\begin{fulltabular}{@{\hspace{\tabcolsep}\extracolsep{\fill}}ccccc} \hline
\multicolumn{2}{c}{10 K}&\multicolumn{2}{c}{30 K} & \\ \cline{1-4}
2$\theta_{\rm obs}$ & $I_{\rm obs}$ & 2$\theta_{\rm obs}$ & $I_{\rm obs}$ & $hkl$\\ \hline
16.27	  &  $19 \pm 8$	&	$-$	&	$-$		&  100 or 10$\overline{1\over 2}$\\
23.38	  &  $78 \pm 8$	&  23.35	&  $71 \pm 8$	&  001\\
33.18	  &  $244 \pm 9$	&  33.17	&  $241 \pm 10$	&  200, 20$\overline{1}$\\
43.03	  &  $138 \pm 8$	&  43.03	&  $140 \pm 9$	&  1$\overline{1}$0\\
44.65	  &  $41 \pm 9$	&  44.63	&  $37 \pm 10$	&  110\\
47.90	  &  $191 \pm 9$	&  47.89	&  $196 \pm 9$	&  201, 20$\overline{2}$, 002, 1$\overline{1}$$\overline{1}$\\
51.52	  &  $102 \pm 7$	&  51.53	&  $100 \pm 9$	&  1$\overline{1}$1\\
60.22	  &  $314 \pm 13$	&  60.22	&  $306 \pm 12$	&  11$\overline{2}$\\
63.27	  &  $548 \pm 19$	&  63.26	&  $566 \pm 19$	&  1$\overline{1}$$\overline{2}$, 3$\overline{1}$$\overline{1}$\\
65.13	  &  1000 	  	&  65.13	&  1000		&  40$\overline{1}$, 31$\overline{1}$, 3$\overline{1}$0\\
68.74	  &  $754 \pm 17$	&  68.73	&  $766 \pm 17$	&  310\\
70.07	  &  $915 \pm 22$	&  70.06	&  $887 \pm 22$	&  1$\overline{1}$2, 40$\overline{2}$, 20$\overline{3}$, 400, 202\\
71.91	  &  $170 \pm 11$	&  71.90	&  $172 \pm 19$	&  31$\overline{2}$, 3$\overline{1}$$\overline{2}$\\
74.78	  &  $100 \pm 11$	&  74.78	&  $92 \pm 11$	&  112, 003\\
76.89	  &  $56 \pm 9$	&  76.85	&  $55 \pm 15$	&  3$\overline{1}$1\\
82.00	  &  $374 \pm 19$	&  81.99	&  $383 \pm 20$	&  11$\overline{3}$, 311\\
84.32	  &  $282 \pm 12$	&  84.30	&  $284 \pm 21$	&  40$\overline{3}$, 401\\
86.26	  &  $73 \pm 21$	&  86.19	&  $56 \pm 23$	&  1$\overline{1}$$\overline{3}$\\
87.26	  &  $663 \pm 26$	&  87.24	&  $672 \pm 25$	&  020, 31$\overline{3}$\\
89.89	  &  $22 \pm 16$	&  89.79	&  $18 \pm 17$	&  3$\overline{1}$$\overline{3}$\\ \hline
\end{fulltabular}
\end{fulltable}

The magnetic reflection appeared at $2\theta = 16.3^\circ$ can be
indexed as either (100) or (10$\overline{1\over 2}$) on the basis of
the original triclinic cell. We cannot make a unique determination of the
magnetic structure because no other magnetic peak was found.  Possible
antiferromagnetic structures which produce either (100) or
(10$\overline{1\over 2}$) magnetic reflection are shown schematically
in Fig.\ \ref{AFstructure}.  The structure (A) corresponds to the
reflection (100) and the structure (B) to the reflection
(10$\overline{1\over 2}$).
\begin{fullfigure}
\centerline{\psfig{figure=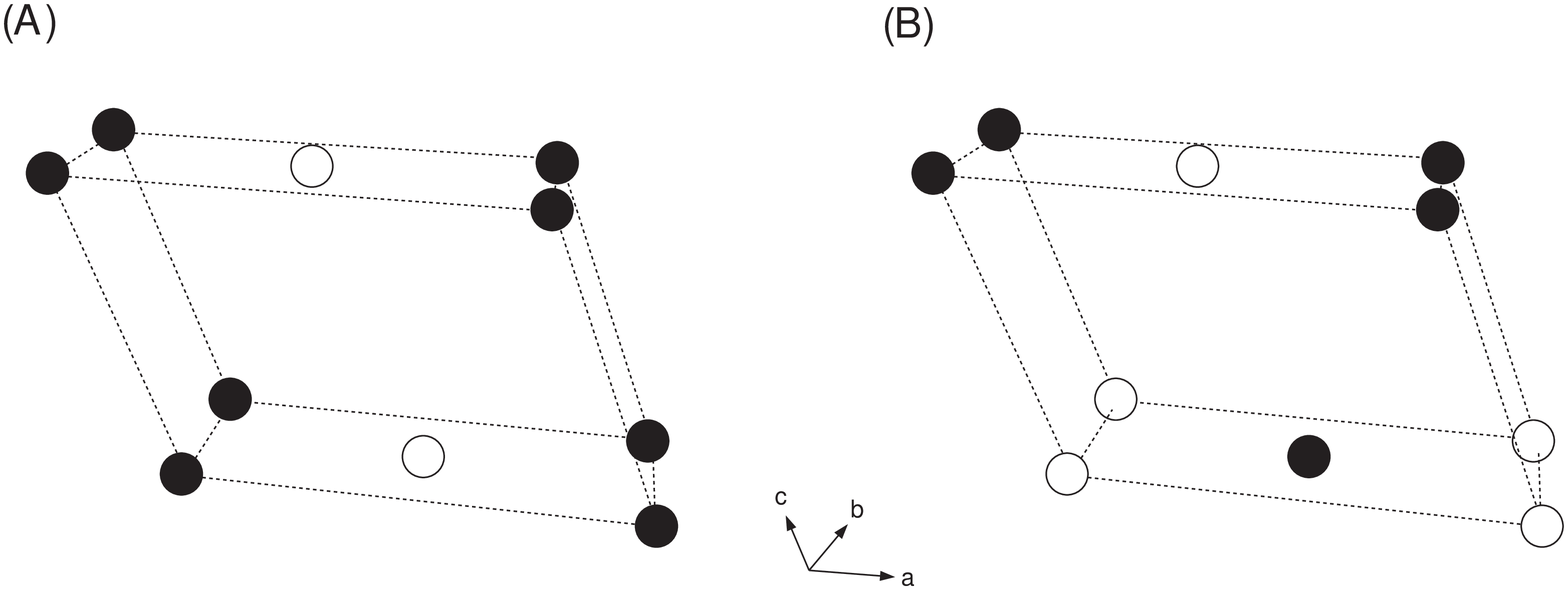,width=15.0cm,angle=0}}
\medskip 
\caption{Two possible magnetic
structures in CuV$_2$O$_6$ which produce (A) (100) and (B)
(10$\overline{1\over 2}$) magnetic reflections, respectively.  Only
the Cu atoms are shown.  Up and down spins are represented by the open
and solid circles, respectively.}
\label{AFstructure}
\end{fullfigure}

It is surprising that in both magnetic structures, the nearest-neighbor
spins align {\it ferromagnetically}.  The antiferromagnetic alignment of the
spins is commonly found along [110] (or [1$\overline{1}$0]), which
is the direction of the next-nearest-neighbor sites of Cu.  It should
be noted that the broad maximum of the susceptibility and the negative
Weiss temperature obtained from the high-temperature region indicate a
dominant antiferromagnetic coupling between Cu$^{2+}$ spins.  This
suggests that the exchange coupling between the next-nearest neighbors
is stronger than that between the nearest neighbors.  The
antiferromagnetic spin chains in CuV$_2$O$_6$ are considered to be along
[110] (or [1$\overline{1}$0]) rather than [010], which is not readily
speculated from the crystal structure.

By symmetry, both (100) and (10$\overline{1\over 2}$) reflections are
purely magnetic in origin, and the intensity $I_{hkl}$ of scattered
neutrons is proportional to the square of the magnetic structure
factor $F_{\rm m}^{hkl}$ as $I_{hkl} \propto \langle q^2 \rangle
|F_{\rm m}^{hkl}|^2$.\cite{marshall71} $F_{\rm m}^{hkl}$ is
proportional to the thermal average of the magnetic moment $\langle
\mu \rangle$ and $\langle q^2 \rangle$ is a geometrical factor depending on
the orientation of the magnetic moment with respect to the scattering
vector.\cite{shirane59} $\langle q^2 \rangle$ is given for the (100)
reflection as
\begin{subequations}
\begin{equation}
\label{q200}
	\langle q^2 \rangle = 1-{d_{100}}^2 {a^{\ast}}^2 \cos^2 \varphi_{a},
\end{equation}

\noindent
and for (10$\overline{1\over 2}$),
\begin{equation}
\label{q201}
	\langle q^2 \rangle = 1-{d_{10\overline{1\over 2}}}^2 (a^{\ast} 
	\cos \varphi_{a} - {c^{\ast}\over 2} \cos \varphi_{c})^2.
\end{equation}
\end{subequations}

\begin{figure}[t]
\centerline{\psfig{figure=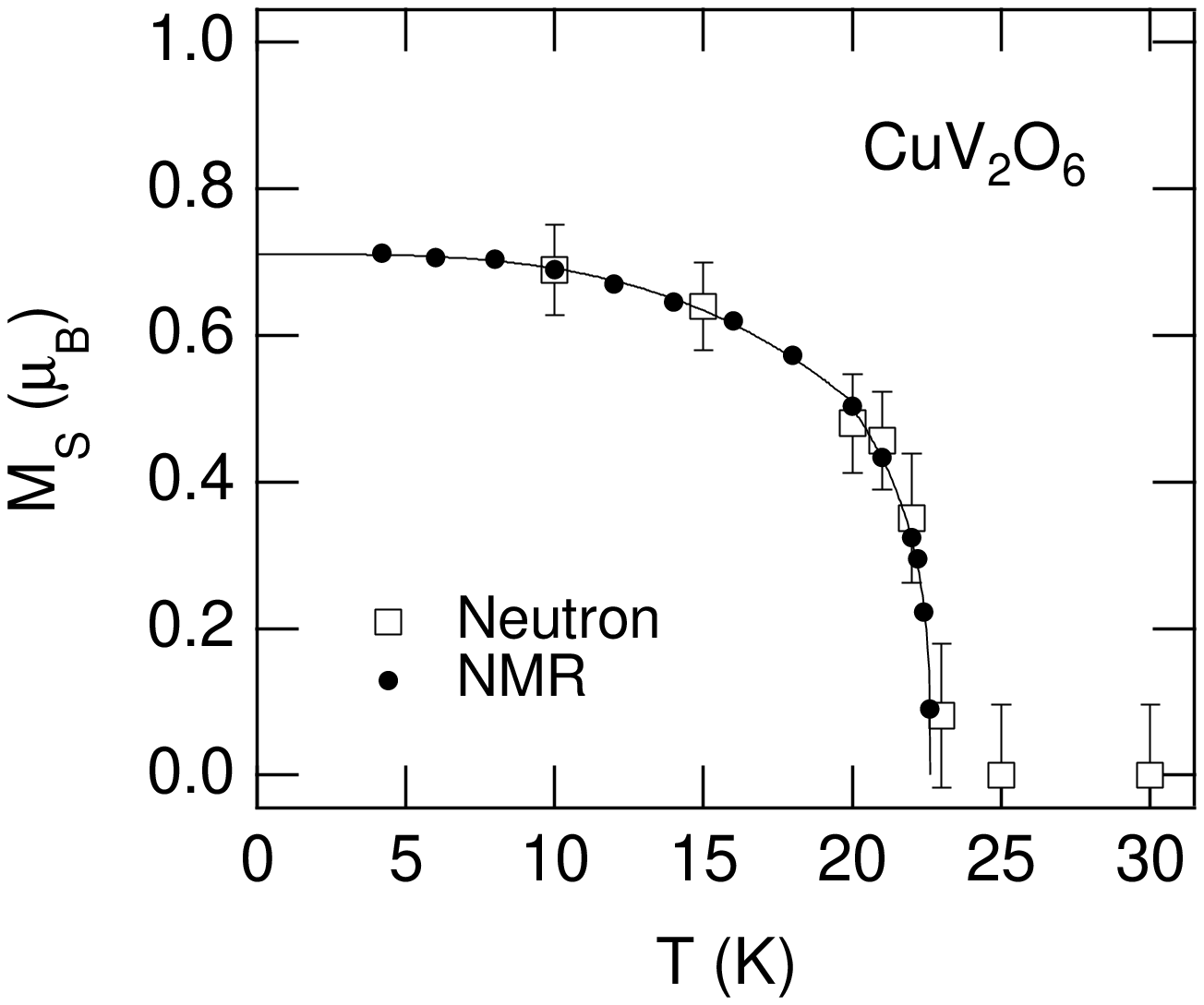,width=9.0cm,angle=0}}
\medskip 
\caption{Temperature dependence of the magnetic moment on Cu atoms
in CuV$_2$O$_6$.  The solid line is a guide to the eyes.  The moment is
calculated for the magnetic structure (A).  The neutron and NMR data
are scaled to match at 10 K, yielding the hyperfine coupling constant
in the antiferromagnetic state $|A_{hf}^{\rm AF}| = 4.3$ kOe/$\mu_{\rm
B}$.  For the structure (B), the vertical axis should be rescaled by multiplying a
factor of 0.965.  }
\label{moment}
\end{figure}

\noindent
Here $d_{hkl}$ is the spacing of $(hkl)$ planes, $a^{\ast}$ and
$c^{\ast}$ are lattice constants in reciprocal space, $\varphi_{a}$ and
$\varphi_{c}$ are angles between the direction of the magnetic moment
and the $a$ and $c$ axes, respectively.  As we have only one magnetic
reflection that is observable, we cannot determine the angles
$\varphi_{a}$, $\varphi_{c}$, and $\langle \mu \rangle$ only from the
neutron diffraction data.  Fortunately, we know from the Cu NMR that
the moment is along the $Z$ principal axis of the EFG tensor at the Cu
site, the direction of which can be evaluated from the symmetry of the
EFG due to surrounding lattice ions.\cite{EFG} The point-charge
calculation of the lattice EFG at the Cu site yields the angles
$\varphi_{a} = 109^\circ$ and $\varphi_{c} = 100^\circ$ between the
$Z$ principal axis of the EFG and the crystalline $a$ and $c$ axes,
respectively. The calculation also yields the angle $\varphi_{b} =
23^\circ$ between the $Z$ principal axis of the EFG and the
$b$ axis, specifying the direction of the magnetic moment with respect
to all the crystalline axes.  For this direction of the magnetic
moment ($\parallel Z$), the intensity of the (010) reflection
at $2\theta = 40.4^\circ$ allowed for the structure (A) is
order-of-magnitude smaller than that of (100).  This is why the
fundamental (010) reflection could not be observed in the present
experiment.  Using the calculated values of $\varphi_{a}$ and
$\varphi_{c}$, and the magnetic form factor of the free Cu$^{2+}$
ion\cite{freeman61} for $F_{\rm m}^{hkl}$, $\langle \mu \rangle$ at 10
K is determined to be $0.69 \pm 0.07~\mu_{\rm B}$ for (100) and $0.67
\pm 0.07~\mu_{\rm B}$ for (10$\overline{1\over 2}$).  The saturation
moment at 0 K is then evaluated to be $0.72 \pm 0.07 \mu_{\rm B}$ and
$0.70 \pm 0.07 \mu_{\rm B}$ for (100) and (10$\overline{1\over 2}$),
respectively, with the aid of the $T$ dependence of the internal
magnetic field at the $^{51}$V site (Fig.\ \ref{moment}).  The
extrapolated values of $\langle \mu \rangle$ at 0 K are slightly
smaller than that expected in the 3D-ordered antiferromagnet with $S =
1/2$,\cite{kubo52} which may be attributed to the low-dimensionality
of the present compound.

\section{Discussion}
\label{sec:discussion}
One of the important points to be discussed here is the
magnetic structure of CuV$_2$O$_6$ revealed by the present
neutron-diffraction measurements.  The antiferromagnetic alignment of
Cu$^{2+}$ spins is found not along [010] (or $b$ axis) which we
thought as the direction of the 1D spin chain, but in the $ab$ plane
probably because of the dominant next-nearest-neighbor coupling.  The
$ab$ plane antiferromagnetism might suggest two-dimensional (2D)
coupling of Cu$^{2+}$ spins rather than the 1D one along either [110]
or [1$\overline{1}$0], and is to be examined as well.  In this
section, we discuss possible exchange paths between Cu$^{2+}$ spins
based on the crystal structure and show that the 1D spin chain is
likely to be along [110].

The magnetic structure of CuV$_2$O$_6$ can be understood by examining
the local symmetry at the Cu site.  Although the atomic site of Cu has
only inversion symmetry, the electronic state of a Cu$^{2+}$ ion may
be derived from the approximate tetragonal symmetry.  In the
CuV$_2$O$_6$ structure, each Cu atom is surrounded by six oxygen atoms
which make a distorted octahedron.  The first and second nearest
neighbors of the six oxygens are the O(1) and O(2) sites forming a
parallelogram with the Cu atom in the center of gravity.  The
distortion of this parallelogram from a square is very small, because
the sides differ by only 1.6 \% (2.774 {\AA} and 2.819 {\AA}) and the
two diagonals O(1)-Cu-O(1) and O(2)-Cu-O(2) make nearly a right angle
($\angle$[O(1)-Cu-O(2)] = 89$^\circ$ at room temperature).  The
octahedron is elongated with two apical oxygens on another O(2) sites,
shared by the neighboring CuO$_6$ octahedra in the $b$
direction as one of the ``planar'' oxygens.  The line O(2)-Cu-O(2) is
canted by 14$^\circ$ from the normal to the plane of the O(1) and O(2)
parallelogram.

If we disregard these distortions and an effect of more distant atoms for
qualitative discussion, the Cu site has tetragonal symmetry in an
elongated octahedral environment, and the Cu$^{2+}$ ion has one hole
in the $d_{x^2-y^2}$ orbital state pointing toward the ``planar''
oxygens on the O(1) and O(2) sites.  The relatively weak exchange between the
nearest-neighbor Cu$^{2+}$ spins can readily be understood because 
there is no relevant orbital to the exchange along [010]
(there are negligible electronic densities pointing toward the ``apical''
oxygens).  It should be noted that the tetragonal symmetry imposes the
EFG at the Cu site to be symmetric about an axis of elongation (or
$\eta = 0$).  The modest asymmetry of the EFG at the Cu site ($\eta =
0.32$) evaluated from the present NMR experiment suggests that the
deviation from tetragonal symmetry is not very large.

\begin{figure}[t]
\centerline{\psfig{figure=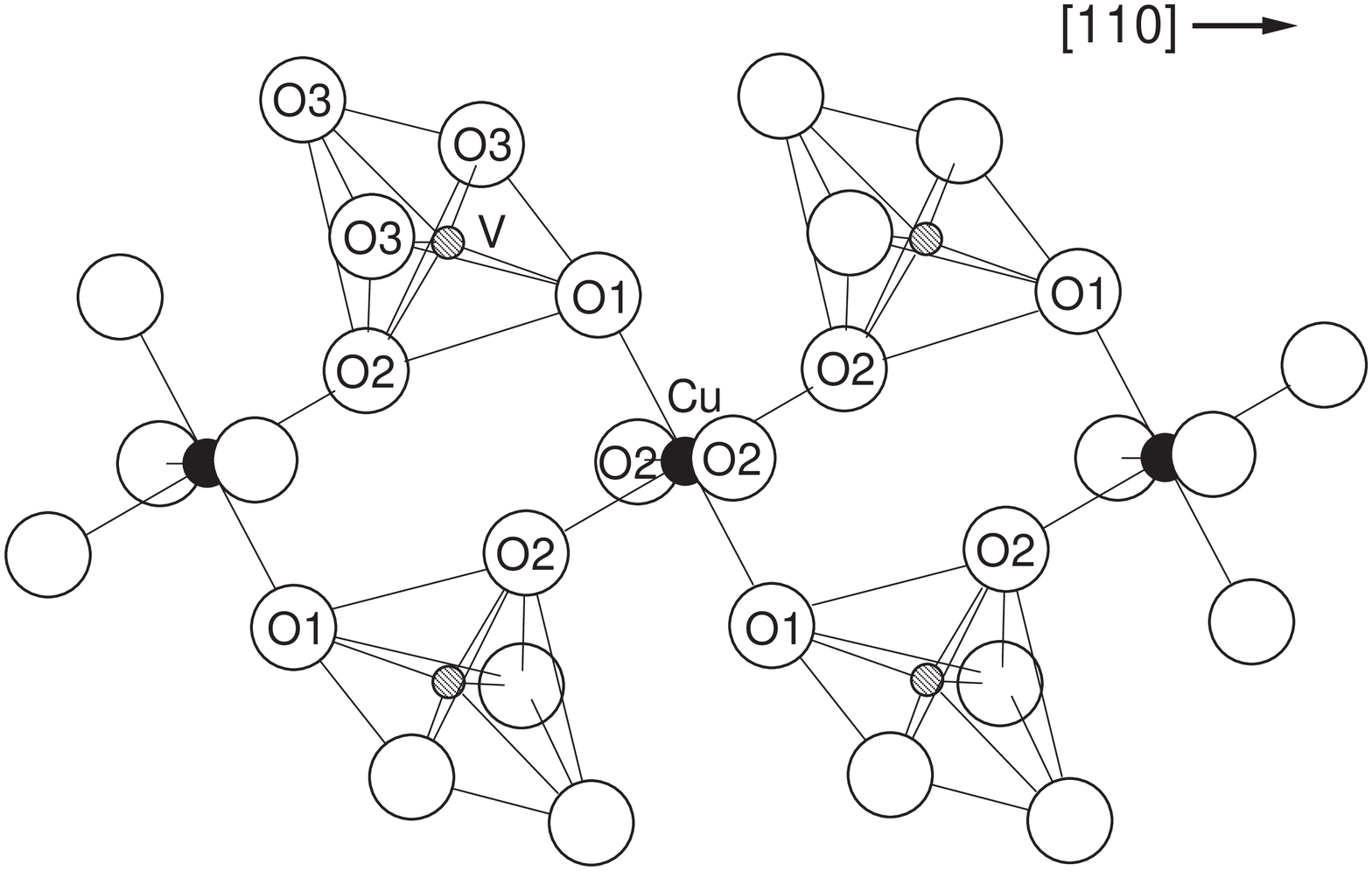,width=8.0cm,angle=0}}
\medskip 
\caption{The suggested Cu chain running along [110] viewed from the apical
oxygens of CuO$_6$ octahedra.  Copper, vanadium and oxygen atoms are
represented by solid, hatched and open circles, respectively.  For
some oxygen atoms, the site indices are explicitly shown.  The CuO$_6$
octahedra are bridged by the two VO$_5$ pyramids successively to make
a one-dimensional chain of Cu atoms along [110].}
\label{chain}
\end{figure}

The $d_{x^2-y^2}$-like character of the electronic state and the linkage
of the CuO$_6$ octahedra favor the exchange along [110], excluding a
possibility of 2D antiferromagnetism in the $ab$ plane with comparable
exchange interactions in both [110] and [1$\overline{1}$0] directions. 
In the [110] direction, the CuO$_6$ octahedra are bridged by two
VO$_5$ pyramids which share the ``planar'' O(1) and O(2) sites of the
CuO$_6$ octahedra as a corner and an apex of the VO$_5$ pyramid,
respectively.  This is shown schematically in Fig.\ \ref{chain}. 
Since the $d_{x^2-y^2}$ orbital is in the plane of the O(1) and O(2)
parallelogram, it may be possible that Cu$^{2+}$ spins have an
exchange coupling in the [110] direction using the bridging VO$_5$
pyramids via the Cu-O-O-Cu path and/or Cu-O-V-O-Cu path.  The
anisotropic hyperfine tensor at the V site not due to the classical
dipolar field seems to be consistent with such exchange paths through
the VO$_5$ pyramids.  On the other hand, there is no such linkage of
the CuO$_6$ octahedra in the [1$\overline{1}$0] direction.  Therefore,
the exchange along [1$\overline{1}$0] is considered to be much weaker
than that along [110].  This suggests that the antiferromagnetic spin
chains in CuV$_2$O$_6$ are along [110].

The distortion from ideal tetragonal symmetry results in the admixture of
the other $3d$ states such as the $d_{z^2}$ orbital pointing toward
the apical oxygens, which enables Cu$^{2+}$ spins to couple
along the $b$ axis.  This is considered to be an origin of long-range
magnetic ordering at non-zero temperature.  The interchain coupling
$J^{\prime}$ may be evaluated using the expression\cite{kondo84}
\begin{equation}
\label{Jinter}
	{T_{\rm N} \over J} = \sqrt{z|J^{\prime}| \over 2J},
\end{equation}

\noindent
where $z$ is the number of nearest-neighbor chains.  Substituting
$T_{\rm N}$ = 22.6 K, $J$ = 36 K and $z$ = 2 into eq.\
(\ref{Jinter}), we obtain $|J^{\prime}|$ = 14 K. Since the interchain
coupling is considered to be ferromagnetic, $J^{\prime}$ may be
negative and thus $J^{\prime} = -14$ K.\cite{fnHex} A similar value of
$J^{\prime}$ has been obtained in the previous report,\cite{vasilev99}
although a different formula is used for the estimation.  A relatively
large value of $|J^{\prime}|$ compared with $J$ indicates poor one
dimensionality of the present system, which may be the origin of a
relatively high ordering temperature.

\section{Conclusion}
\label{conclusion}
We have measured the susceptibility, NMR and powder neutron diffraction in
an $S = 1/2$ quasi one-dimensional magnet CuV$_2$O$_6$.  It was confirmed that
CuV$_2$O$_6$ exhibits three-dimensional antiferromagnetic ordering at
the N\'{e}el temperature of 22.6 K, which is not far below the rounded
maximum of the susceptibility observed around 48 K. The two possible
antiferromagnetic structures were suggested and the saturation moments were
evaluated to be 0.72 and 0.70 $\mu_{\rm B}$ for the respective
magnetic structures.  The intra- and interchain couplings were also
estimated to be 36 and $-$14 K, respectively.

The magnetic structure of CuV$_2$O$_6$ indicates that the exchange coupling
between Cu$^{2+}$ spins is stronger in the direction of next-nearest
neighbors than in the direction of nearest neighbors.  The origin of
such exchange couplings was argued based on the local symmetry at the
Cu site and the linkage of the CuO$_6$ octahedra in the crystal
structure.  It is suggested that the $d_{x^2-y^2}$-like character of the
electronic state inferred from symmetry is responsible for the
relatively weak exchange between the nearest-neighbors spins along the
$b$ axis.  A dominant antiferromagnetic exchange compatible with the
$d_{x^2-y^2}$-like orbital character may be the one acting via the
Cu-O-O-Cu interaction paths, suggesting one-dimensional chains of
Cu$^{2+}$ spins along [110].  Because of such distant exchange paths
for the intrachain coupling, CuV$_2$O$_6$ has a relatively large ratio of
the interchain coupling to the intrachain one, which makes the N\'{e}el
temperature of this compound relatively high.

\appendix
\section{Field-swept NMR spectrum in the antiferromagnetic state}
In this appendix, we will derive a powder pattern of the field-swept NMR
spectrum in the antiferromagnetic state.  In the presence of both
internal magnetic field ${\mib H_{\rm n}}$ and external magnetic field ${\mib
H}$, the resonance condition for a given nucleus is
expressed as
\begin{equation}
\label{condvector}
	\omega /\gamma = |{\mib H} + {\mib H_{\rm n}}|.
\end{equation}

\noindent
This can be written alternatively as
\begin{equation}
\label{condscalar}
	\omega^2/\gamma^2 = H^2 + H_{\rm n}^2 + 2HH_{\rm n} \cos \theta,
\end{equation}

\noindent
where $H = |{\mib H}|$, $H_{\rm n} = |{\mib H_{\rm n}}|$ and $\theta$
is the angle between ${\mib H}$ and ${\mib H_{\rm n}}$.  For a powder
sample, the direction of ${\mib H}$ is randomly distributed with respect to
${\mib H_{\rm n}}$. The probability of ${\mib H}$ and ${\mib
H_{\rm n}}$ making an angle between $\theta$ and $\theta + \Delta \theta$
is proportional to the solid angle $\Delta \Omega \propto \sin \theta
\Delta \theta$.  The number of nuclei $\Delta N \equiv f(H) \Delta H$ such
that the resonance field has a value between $H$ and $H +\Delta H$ is
proportional to $\Delta \Omega$ or $f(H)\Delta H \propto \sin \theta
\Delta \theta$, giving the NMR line shape $f(H)$;
\begin{equation}
\label{spectrum1}
	f(H) \propto \sin \theta {d\theta \over dH}.
\end{equation}

\noindent
$d\theta /dH$ can be calculated by differentiating eq.\ (\ref{condscalar}). 
Substituting $d\theta /dH$ into eq.\ (\ref{spectrum1}) yields
\begin{equation}
\label{spectrum2}
	f(H) \propto {{H^2-H_n^2+\omega^2 /\gamma^2}\over {H_n H^2}}.
\end{equation}

\noindent
$f(H)$ is zero outside the region $\omega /\gamma - H_{\rm n} \leq H
\leq \omega /\gamma + H_{\rm n}$ because $0 \leq \theta \leq \pi$. 
The spectrum is therefore cutoff at fields $\omega /\gamma \pm H_{\rm
n}$.  In fact, an inhomogeneous distribution of $H_{\rm n}$ present in
a real crystal broadens the spectrum\ (\ref{spectrum2}), so that the sharp
edges at the cutoff fields are obscured.  The spectrum to be observed
in the presence of an inhomogeneous broadening is now given by the
convolution of eq.\ (\ref{spectrum2}) and the distribution function
$g(H^{\prime})$ for $H_{\rm n}$ as
\begin{equation}
\label{broadened}
	F(H) = \int f(H-H^{\prime}) g(H^{\prime}) dH^{\prime}.
\end{equation}

\begin{figure}[t]
\centerline{\psfig{figure=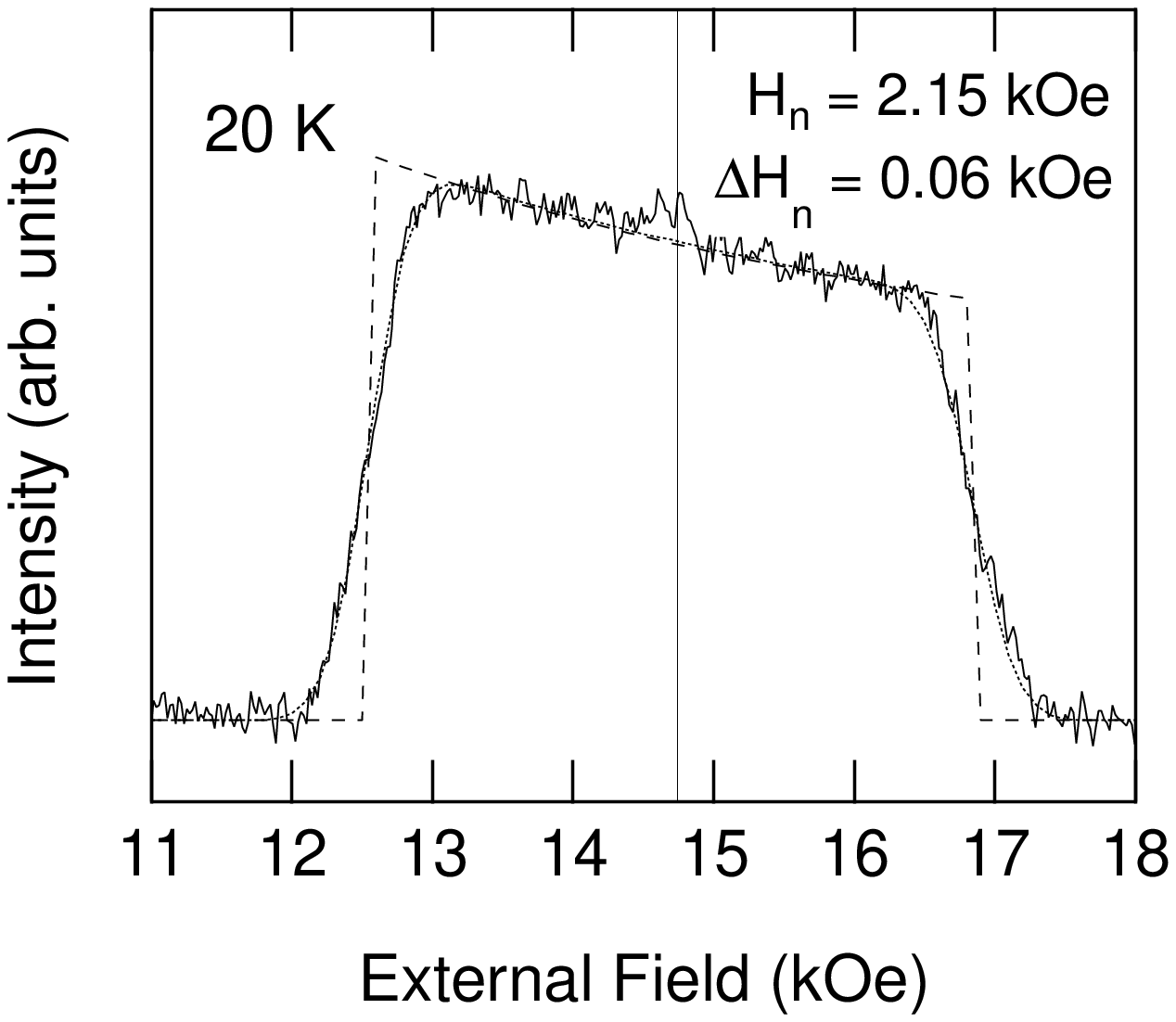,width=8.5cm,angle=0}}
\medskip 
\caption{Comparison of the calculated powder patterns and the observed
$^{51}$V NMR spectrum in CuV$_2$O$_6$ at 16.5 MHz and 20 K. The dashed
line is the spectrum with no inhomogeneous broadening, eq.\
(\ref{spectrum2}), showing sharp edges at $\omega /\gamma \pm H_{\rm
n}$.  $\omega /\gamma$ corresponds to the zero-shift position for
$^{51}$V nuclei which is represented by the vertical line.  The edges
are equally spaced about $\omega /\gamma$ by $H_{\rm n}$ giving the
spectral width of $2H_{\rm n}$.  The dotted line is a fit of the
experimental data to eq.\ (\ref{broadened}) for which the distribution
function $g(H^{\prime}) \propto \exp ({-{1\over 2}{{(H^{\prime}-H_{\rm
n})^2} \over {\Delta H_{\rm n}}^2}})$ is assumed.  The least-squares
fit gives $H_{\rm n} = 2.15$ kOe and $\Delta H_{\rm n}=0.06$ kOe at 20
K.}
\label{51VAF}
\end{figure}

\noindent
As one usually assumes, we used a gaussian distribution function for
$g(H^{\prime})$ to reproduce $^{51}$V NMR spectra in the antiferromagnetic
state.  In Fig.\ \ref{51VAF} we showed the results of the fit and
an ``ideal'' spectrum with no inhomogeneous broadening (corresponding to the
case $g(H^{\prime}) = \delta(H^{\prime})$).  From the least-squares
fit of the data, we obtained $H_{\rm n} = 2.15$ kOe at 20 K. The standard
deviation of $H_{\rm n}$ was 0.06 kOe at 20 K which was taken to be
the error in evaluating $H_{\rm n}$.



\begin{thebibliography}{99}
\bibitem{fnSG} In the notation of ref.\ \citen{calvo73}, the space
group of CuV$_2$O$_6$ is $C\overline{1}$.  This is an unconventional setting
of $P\overline{1}$ with symmetry transforms: $x,y,z$; $-x,-y,-z$;
${1\over 2}+x,{1\over 2}+y,{1\over 2}+z$; ${1\over 2}-x,{1\over
2}-y,{1\over 2}-z$, yielding a $C$-centered triclinic cell.
\bibitem{nQ65} The quadrupole frequency $^{65}\nu_{\rm Q}$ for $^{65}$Cu
is calculated from the isotopic ratio of quadrupole moments
$^{65}Q/^{63}Q$ = 0.925 as ${^{65}\nu_{\rm Q}} = {^{63}\nu_{\rm
Q}}({^{65}Q}/{^{63}Q})$.
\bibitem{EFG} There is also a contribution of the valence electron to
the EFG at the Cu site which is known to affect significantly the magnitude of
$\nu_{\rm Q}$.  However, it is expected that the symmetry of the EFG
represented by $\eta$ is not affected so much because the electronic
orbital of the valence electron reflects the local symmetry of the
site under consideration.  The fact that the observed asymmetry
$\eta_{\rm obs}=0.32$ at the Cu site agrees well with the asymmetry of the
lattice EFG, $\eta_{\rm latt}=0.33$, evaluated by the point-charge model seems
to support the above expectation.
\bibitem{fnHex} We take the positive (negative) sign of the exchange constant for
the antiferromagnetic (ferromagnetic) coupling following the
convention for the 1D Heisenberg spin chain with
nearest-neighbor coupling ${\cal H} = J \sum_{i} {\mib S}_{i} \cdot
{\mib S}_{i+1}$.
\end{thebibliography}
\end{document}